\documentclass[12pt,preprint]{emulateapj}

\usepackage{color}
\usepackage{ascmac}

\slugcomment{Accepted: July 31, 2012}

\shorttitle{Grain Growth in Very Metal-Poor Clouds}
\shortauthors{Nozawa, Kozasa, \& Nomoto}

\begin{document}

\title{Can the Growth of Dust Grains in Low-Metallicity Star-Forming \\
Clouds Affect the Formation of Metal-Poor Low-Mass Stars?}

\author{
Takaya Nozawa\altaffilmark{1},
Takashi Kozasa\altaffilmark{2},
and Ken'ichi Nomoto\altaffilmark{1}}

\altaffiltext{1}{Kavli Institute for the Physics and Mathematics of the
Universe (WPI), Todai Institutes for Advanced Study, The University of 
Tokyo, Kashiwa, Chiba 277-8583, Japan; takaya.nozawa@ipmu.jp}
\altaffiltext{2}{Department of Cosmosciences, Graduate School 
of Science, Hokkaido University, Sapporo 060-0810, Japan}

\begin{abstract}

The discovery of a low-mass star with such low metallicity as 
$\le 4.5 \times 10^{-5}$ $Z_\odot$ reveals the critical role of dust
in the formation of extremely metal-poor stars.
In this paper we explore the effect of the growth of dust grains through 
accretion of gaseous refractory elements in very low-metallicity 
pre-stellar cores on the cloud fragmentation induced by the dust 
emission cooling. 
Employing a simple model of grain growth in a gravitationally 
collapsing gas, we show that Fe and Si grains can grow efficiently 
at hydrogen densities of $\simeq 10^{10}$--$10^{14}$ cm$^{-3}$ in the
clouds with metal abundances of $-5 \la$ [Fe, Si/H] $\la -3$.
The critical metal number abundances, above which the grain growth 
could induce the fragmentation of the gas clouds, are estimated to be 
$A_{\rm crit} \simeq$ $10^{-9}$--$10^{-8}$, unless the initial grain 
radius is too large ($\ga$ 1 $\mu$m) or the sticking probability is 
too small ($\la$ 0.01).
We find that even if the initial dust-to-gas mass ratio is well below 
the minimum value required for the dust-induced fragmentation, the 
grain growth increases the dust mass high enough to cause the gas 
fragmentation into sub-solar mass clumps. 
We suggest that as long as the critical metal abundance is satisfied,
the grain growth could play an important role in the formation of 
low-mass stars with metallicity as low as $10^{-5}$ $Z_\odot$.

\end{abstract}


\section{Introduction}

Dust grains in the early universe are considered to be important 
agents to trigger the formation of low-mass stars in metal-poor 
environments (Omukai 2000; Schneider et al.\ 2003);
the cooling of the gas through thermal emission of dust makes
collapsing dense cores gravitationally unstable, leading to the
fragmentation into multiple sub-solar mass clumps at gas densities of 
$10^{12}$--$10^{14}$ cm$^{-3}$ 
(Tsuribe \& Omukai 2006; Dopcke et al.\ 2011).
This scenario has been recently supported by the discovery of a 
Galactic low-mass star, SDSS J102915+172927 (Caffau et al.\ 2011),
whose metal content is too low ($Z \le 4.5 \times 10^{-5}$ $Z_\odot$) 
to induce the fragmentation of star-forming clouds by metal-line cooling
(see Klessen et al.\ 2012; Schneider et al.\ 2012b for details).

The condition that realizes the dust-induced fragmentation depends 
on the amount of dust grains as well as their size distribution in 
pre-stellar clouds (Omukai et al.\ 2005; Schneider et al.\ 2006, 2012a).
Schneider et al.\ (2012a) found that the formation condition of the 
low-mass fragments obtained by the numerical simulations is fully 
described in terms of the product of dust-to-gas mass ratio 
$\mathcal{D}$ and geometrical cross section per unit dust mass 
$\mathcal{S}$ as follows;
\begin{eqnarray}
\mathcal{S D} > 1.4 \times 10^{-3} ~ {\rm cm}^2 ~ {\rm g}^{-1}
\left( \frac{T_{\rm gas}}{10^3 ~ {\rm K}} \right)^{-\frac{1}{2}}
\left( \frac{c_{\rm H}}{10^{12} ~ {\rm cm}^{-3}} 
\right)^{-\frac{1}{2}},
\end{eqnarray}
where $T_{\rm gas}$ is the temperature of the gas, and $c_{\rm H}$ is 
the hydrogen number density.
Treating self-consistently the dust formation in the ejecta of 
supernovae (SNe) and the subsequent destruction of the dust by the 
reverse shocks, Schneider et al.\ (2012a) argued that the condition 
could not be satisfied in the collapsing star-forming clouds enriched 
with metals and dust from the first SNe if a majority of grain 
formed in the SN ejecta are destroyed by the reverse shock
(see also Schneider et al.\ 2012b).
However, it could be possible that the accretion of gaseous refractory 
elements released from dust grains in the shocked gas onto the 
surfaces of the SN dust surviving in star-forming clouds changes the 
mass and size distribution of the dust, and thus affects the thermal 
evolution of the collapsing cores.\footnote{Hirashita \& Omukai (2009) 
examined the coagulation of dust in collapsing clouds with a variety 
of metallicity.
They found that the dust coagulation can proceed even at metallicity 
as low as $10^{-6}$ $Z_\odot$ but does not have any impact on the 
thermal evolution of the star-forming clouds.}

In this paper, we investigate the feasibility of grain growth in 
low-metallicity star-forming clouds to explore whether the grain 
growth can facilitate the formation of metal-poor low-mass stars.
In Section 2, we describe the model of grain growth in collapsing 
dense clouds, and present the results of the calculations in Section 3.
In Section 4, we estimate the critical metal abundances above which the 
grain growth could encourage the gas fragmentation into sub-solar mass 
clumps, and discuss the corresponding dust-to-gas mass ratio and total 
metallicity.  
The conclusion is given in Section 5.

\section{Model of Grain Growth in Metal-Poor Star-Forming Clouds}

We consider the growth of dust grains in collapsing clouds that have 
been enriched with metals and dust grains produced by very early 
generation of SNe.
Dust formation calculations by Nozawa et al.\ (2003) showed that 
various grain species condense in the unmixed ejecta of Population 
III SNe, and that Fe, Si, and C grains have relatively large average 
radii ($\ga$ 0.01 $\mu$m). 
Based on their dust models, Nozawa et al.\ (2007) investigated the 
evolution of dust in the hot gas swept up by the SN shocks and found 
that most of such large Fe, Si and C grains can survive the destruction 
by the reverse shock to be predominantly injected into the early 
interstellar medium.

Being motivated by these studies, and to simplify the chemistry, we 
consider that Fe and Si grains composed of only one element grow 
through accretion of Fe and Si atoms in the gas phase, 
respectively.\footnote{The growth of C grains may not be expected in 
dense clouds. 
This is because at high gas densities considered in this paper 
($c_{\rm H} \ge 10^8$ cm$^{-3}$), all C atoms can be bounded in CO 
molecules in gas clouds with metallicity scaled by the solar abundance
(Omukai et al.\ 2005).}
We assume that dust grains are spheres, all of which have 
a single initial radius, although they might be expected to have the 
distribution of grain sizes.
Suppose that the number density of a given refractory element $i$ 
at a time $t_0$ is $c_{i,0} = c_i(t_0) = A_i c_{\rm{H},0}$, where
$A_i$ is the number abundance of the element $i$ relative to 
hydrogen.
In order to specify the number abundance of pre-existing seed grains, 
we introduce a parameter $f_{i,0}$ ($>$0), which is defined as the 
number fraction of the elements $i$ originally locked in dust grains 
(i.e., condensation efficiency at $t_0$).
Then, the number density of the dust whose initial radius is
$r_{i,0}$ is described as
$n_{i,0}^{\rm dust} = f_{i,0} c_{i,0} (a_{i,0} / r_{i,0})^3$, 
where $a_{i,0}$ is the hypothetical radius of an atom in the dust 
phase.

The time evolution of the number density $c_i(t)$ of an element $i$ 
in the gas clouds collapsing with the timescale of free-fall is 
given by
\begin{eqnarray}
c_i(t) = c_{i,0} \left( 1 - \frac{t}{2 \tau_0^{\rm ff}} 
\right)^{-2},
\end{eqnarray}
where $t$ is the elapsed time from $t_0$,
$\tau_0^{\rm ff} = (3 \pi /32 G \mu m_{\rm H} c_{{\rm H},0})^{1/2}$ 
is the free-fall time at the density $c_{{\rm H},0}$ with the 
gravitational constant $G$, the mean molecular weight $\mu$, and the 
mass of a hydrogen atom $m_{\rm H}$.
Once the grain growth activates, the gaseous atoms are consumed, and
the number density $c_i^{\rm gas}(t)$ of an element $i$ in the gas 
phase at the time $t$ can be written as
\begin{eqnarray}
c_i^{\rm gas}(t) = c_i(t) \{ 1 - f_{i,0} 
\left[ r_i(t) / r_{i,0} \right]^3 \},
\end{eqnarray}
with $r_i(t)$ being the radius of the $i$-th grain species at $t$.
Equation (3) is reduced to
\begin{eqnarray}
f_i(t) = 1 - Y_i(t) = f_{i,0} X_i^3(t),
\end{eqnarray}
where $f_i(t)$ is the condensation efficiency at $t$, 
$Y_i(t) = c_i^{\rm gas}(t) / c_i(t)$ represents the depletion of the 
gaseous atoms due to grain growth, and $X_i(t) = r_i(t) / r_{i,0}$.

In the dense clouds where almost all gaseous atoms are neutral, 
the growth rate of grain radius is given by
\begin{eqnarray}
\frac{dr_i}{dt} = s_i \left( \frac{4 \pi}{3} a_{i,0}^3 \right) 
\left( \frac{k T_{\rm gas}}{2 \pi m_i} \right)^{\frac{1}{2}} 
c_i^{\rm gas}(t) \left( 1 - \frac{1}{S_i} 
\sqrt{ \frac{T_{\rm dust}}{T_{\rm gas}} } \right),
\end{eqnarray}
where $s_i$ is the sticking probability of the gaseous element $i$ 
incident onto grains, $k$ is the Boltzmann constant, and $m_i$ is the 
mass of the element $i$.
The gas temperature $T_{\rm gas}$ is assumed to be constant during the 
evolution of clouds in this study.
The supersaturation ratio $S_i$ is a function of the dust temperature 
$T_{\rm dust}$.
Since $T_{\rm dust} / T_{\rm gas} \ll 1$ and $S_i \gg 1$ under the 
condition considered here (Dopcke et al.\ 2011), Equation (5) is 
reduced to
\begin{eqnarray}
\frac{dX_i}{dt} = \frac{ Y_i(t)}{\tau_{i,0}^{\rm gg}}
\left( 1 - \frac{t}{2 \tau_0^{\rm ff}} \right)^{-2},
\end{eqnarray}
by introducing $(\tau_{i,0}^{\rm gg})^{-1}$ = $s_i 4 \pi a_{i,0}^3 
\left( k T_{\rm gas} / 2 \pi m_i \right)^{1/2} 
A_i c_{{\rm H},0} / 3 r_{i,0}$.
Then, integration of Equation (6) leads to the ratio of grain radius 
to the initial one $X_i(t)$;
\begin{eqnarray}
X_i(t) = 1 + \frac{2 \tau_0^{\rm ff}}{\tau_{i,0}^{\rm gg}} 
\int_0^u \frac{Y(u')}{\left( 1 - u' \right)^2} du',
\end{eqnarray}
with $u = t /2 \tau_0^{\rm ff}$.

In principle, by solving Equations (4) and (7) for a given set of
$f_{i,0}$, $r_{i,0}$, and $A_i$, we can calculate the time evolution 
of $f_i(t)$ and $r_i(t) = r_{i,0} X_i(t)$.
The values of the other parameters necessary for the calculations
are summarized in Table 1.
As is shown later, for the metal abundances considered in this paper, 
the grain growth operates at high gas densities of
$c_{\rm H} \ga 10^{10}$ cm$^{-3}$, where the gas temperature is expected
to be in the range of 500--2000 K (Dopcke et al.\ 2011).
Thus, the calculations are started from $c_{{\rm H},0} = 10^8$ 
cm$^{-3}$, with $T_{\rm gas} = 10^3$ K.
Note that the results of calculations are not sensitive to 
$T_{\rm gas}$ as long as the above range of $T_{\rm gas}$ is considered.

\begin{deluxetable}{ll}
\tablewidth{0pt}
\tablecaption{Numerical Values Used in the Calculations}
\tablehead{ 
\colhead{Numerical Values} & \colhead{Explanation of Symbols}
}
\startdata
$s = 1$                  & sticking probability \\
$T_{\rm gas} = 10^3$ K   & gas temperature \\
$\mu = 2.18$             & mean molecular weight \\
$a_{{\rm Fe},0} = 1.441$~\AA & radius of a Fe atom in the solid 
phase\tablenotemark{a} \\
$a_{{\rm Si},0} = 1.684$~\AA & radius of a Si atom in the solid 
phase\tablenotemark{a} \\
$m_{\rm Fe} = 56 m_{\rm H}$ & mass of a Fe atom \\
$m_{\rm Si} = 28 m_{\rm H}$ & mass of a Si atom \\
$A_{{\rm Fe},\odot} = 3.26 \times 10^{-5}$ & solar abundance of Fe
relative to H\tablenotemark{b} \\
$A_{{\rm Si},\odot} = 3.58 \times 10^{-5}$ & solar abundance of Si
relative to H\tablenotemark{b} \\
$\rho_{\rm Fe} = 7.90$ g cm$^{-3}$ & bulk density of 
Fe\tablenotemark{c} \\
$\rho_{\rm Si} = 2.32$ g cm$^{-3}$ & bulk density of 
Si\tablenotemark{c}
\enddata
\tablenotetext{a}{Nozawa et al.\ (2003)}
\tablenotetext{b}{Anders \& Grevesse (1989)}
\tablenotetext{c}{Nozawa et al.\ (2006)}
\end{deluxetable}

\section{Results of calculations of grain growth}

Figure 1 depicts the growth of Fe grains with $f_{{\rm Fe},0} = 0.1$ 
and $r_{{\rm Fe},0} = 0.01$ $\mu$m;
the time evolutions of the condensation efficiency $f_{\rm Fe}(t)$ 
and grain radius $r_{\rm Fe}(t)$ versus hydrogen number density 
$c_{\rm H}(t)$ for [Fe/H] = $-5$, $-4$, and $-3$, which correspond 
to $A_{\rm Fe} = 3.26 \times 10^{-10}$, $3.26 \times 10^{-9}$, and 
$3.26 \times 10^{-8}$, respectively, with the solar abundance by 
Anders \& Grevesse (1989).
We can see that the grain growth activates efficiently even in the 
gas clouds with [Fe/H] = $-5$, and the grain radius finally reaches 
a constant value $r_{{\rm Fe},0} ( 1 / f_{{\rm Fe},0} )^{1/3}$ by 
consuming up all gaseous Fe atoms.
However, the gas density at which a considerable fraction 
($f_{\rm Fe} \sim 0.5$) of Fe atoms is locked up in dust grains 
is higher for a lower Fe abundance; 
$c_{\rm H} \simeq 10^{10}$, $10^{12}$, and $10^{14}$ cm$^{-3}$ for 
[Fe/H] = $-3$, $-4$, and $-5$, respectively. 
Also, the gas density at which $f_{\rm Fe}$ reaches $\sim$0.5 is 
two orders of magnitude higher (lower) for $r_{{\rm Fe},0} = 0.1$ 
(0.001) $\mu$m 
than that for $r_{{\rm Fe},0} = 0.01$ $\mu$m, though not presented 
in the figure.
These behaviors of grain growth in the collapsing gas clouds can be
seen from Equation (9) (see below);
for given values of $f_{i,0}$ and $f_{i,*}$,
$c_{\rm H}(t) \propto (r_{i,0} / A_i)^2$ if
$c_{\rm H}(t) / c_{{\rm H},0} \gg 1$.
Hence, for a fixed $f_{i,0}$, one order of magnitude higher 
$r_{i,0}$ or one order of magnitude lower $A_i$ is compensated with 
two orders of magnitude higher $c_{\rm H}$.

\begin{figure}
\epsscale{1.1}
\plotone{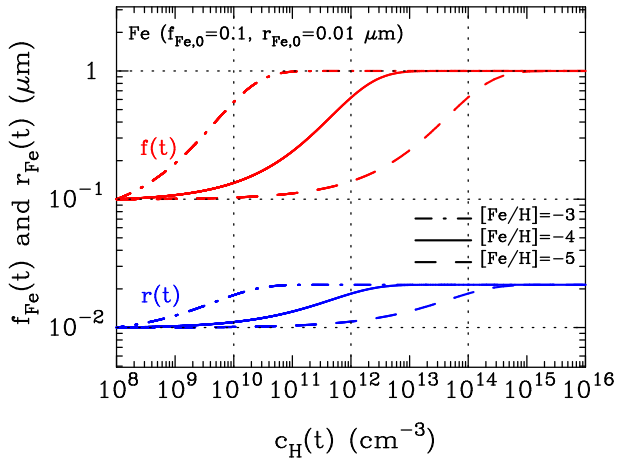}
\caption{
 Time evolutions of the condensation efficiency $f_{\rm Fe}(t)$ (red) 
 and grain radius $r_{\rm Fe}(t)$ (blue) through the growth of Fe 
 grains with the initial dust abundance $f_{{\rm Fe},0} = 0.1$ and the 
 initial grain radius $r_{{\rm Fe},0} = 0.01$ $\mu$m as a function of 
 hydrogen number density $c_{\rm H}(t)$.
 Dot-dashed, solid, and dashed lines depict the results for the 
 collapsing clouds with the Fe abundances of [Fe/H] = $-3$, $-4$, and 
 $-5$, respectively.
\label{fig1}}
\end{figure}

Figure 2 shows the time evolutions of $f_{\rm Si}(t)$ and 
$r_{\rm Si}(t)$ of Si grains with 
$r_{{\rm Si},0} = 0.01$ $\mu$m for different initial dust abundances 
of $f_{{\rm Si},0} = 0.1$, 0.01, and 0.001.
Here, the total abundance of Si atoms is set to be [Si/H] = $-4$ 
($A_{\rm Si} = 3.58 \times 10^{-9}$).
For $f_{i,0} = 0.1$, the growth of Si grains proceeds somewhat 
earlier than that of Fe grains for [Fe/H] $= -4$, and the condensation 
efficiency increases to $f_{\rm Si} = 0.5$ at 
$c_{\rm H} \simeq 10^{11}$ cm$^{-3}$.
In the cases of lower $f_{{\rm Si},0}$, higher gas densities are 
needed for achieving some level of the condensation efficiency, 
although the final grain radii are larger when all Si atoms are tied up
in dust grains.

\begin{figure}
\epsscale{1.1}
\plotone{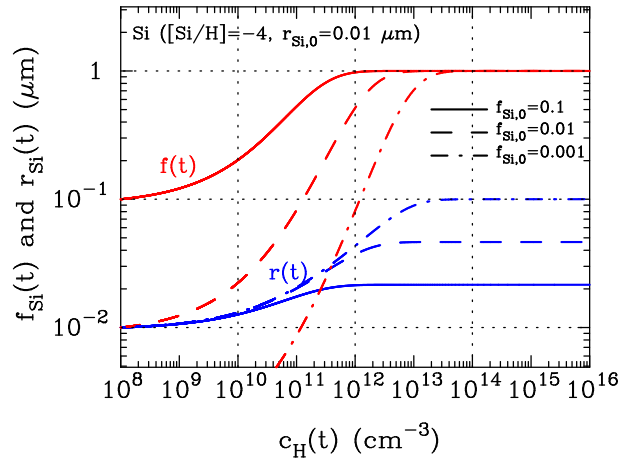}
\caption{
 Time evolutions of the condensation efficiency $f_{\rm Si}(t)$ (red)
 and grain radius $r_{\rm Si}(t)$ (blue) through the growth of Si 
 grains of $r_{{\rm Si},0} = 0.01$ $\mu$m in the collapsing clouds 
 with the Si abundance of [Si/H] = $-4$.
 Solid, dashed, and dot-dashed lines represent the results for the 
 initial dust abundances of $f_{{\rm Si},0} = 0.1$, 0.01, and 0.001, 
 respectively.
\label{fig2}}
\end{figure}

\section{Discussion: Critical Metal Abundances}

As shown in the last section, the grain growth can activate even in 
metal-poor star-forming clouds whose metallicity is only [Fe, Si/H] 
$\simeq$ $-5$.
However, in order that the grain growth affects the thermal evolution 
of collapsing cores, it must become effective before the cloud density 
increases to $c_{\rm H} = 10^{12}$--$10^{14}$ cm$^{-3}$, where the 
optical depth becomes high enough to suppress dust emission cooling,
and as a result the gas fragmentation is expected to occur
(e.g., Schneider et al.\ 2012a).

The metal abundance above which the grain growth becomes important 
can be estimated by requiring that a certain fraction $f_{i,*}$ of the 
element $i$ should be locked up in dust grains at a given hydrogen 
number density $c_{{\rm H},*}$.
Noting that $Y_i(t) = 1 - f_{i,0} X_i^3(t) = 1 - f_i(t)$,
Equation (7) can be rewritten as
\begin{eqnarray}
K_i(f_{i,0}, f_{i,*}) 
&=& \int_1^{X_{i,*}} \frac{dX_i}{1 - f_{i,0} X_i^3} \nonumber  \\
&=& \frac{2 \tau_0^{\rm ff}}{\tau_{i,0}^{\rm gg}}
\int_0^{u_*} \frac{du'}{\left( 1 - u' \right)^2} 
= \frac{2 \tau_0^{\rm ff}}{\tau_{i,0}^{\rm gg}}
\frac{u_*}{ 1 - u_* },
\end{eqnarray}
where $X_{i,*} = (f_{i,*} / f_{i,0})^{1/3}$ and 
$u_* = 1 - (c_{{\rm H},0} / c_{{\rm H},*})^{1/2}$.
From this equation, we can derive the critical metal abundances above 
which the grain growth could facilitate the fragmentation of the 
clouds as follows
\begin{eqnarray}
A_{i,\rm{crit}} = (1.0-2.5) \times 10^{-9} \ K_i
\left( \frac{r_{i,0}}{0.01~\mu{\rm m}} \right)
\left( \frac{10^{12}~{\rm cm}^{-3}}{c_{{\rm H},*}} \right)^{\frac{1}{2}},
\end{eqnarray}
using $c_{{\rm H},0}/c_{{\rm H},*} \ll 1$.
The numerical factor 2.5 (1.0) corresponds to Fe (Si) grains.
The function $K_i$ increases with increasing $f_{i,*}$ and/or 
decreasing $f_{i,0}$;
for $0.2 \le f_{i,*} \le 0.8$, $K_i =$ 0.3--2.4 (5.2--15) at 
$f_{i,0} = 0.1$ ($f_{i,0} = 0.001$).

Figure 3 presents the critical abundances of Fe and Si in the form of 
[X/H] versus $f_{i,0}$ for $r_{i,0} = 0.01$ $\mu$m;  
the dot-dashed, solid, and dashed lines give the abundances necessary 
for the condensation efficiency $f_{i,*}$ to reach 0.8, 0.5, and 0.25, 
respectively, at $c_{\rm H,*} = 10^{12}$ cm$^{-3}$.
As expected, higher metal abundances are needed for attaining higher 
$f_{i,*}$ and/or for lower $f_{i,0}$.
For the case of $f_{i,*} = 0.5$, the critical abundances of Fe and Si 
spans the ranges of $-4.11 \le$ [Fe/H] $\le -3.19$ and 
$-4.54 \le$ [Si/H] $\le -3.62$, respectively, for the range of 
$0.1 \ge f_{i,0} \ge 0.001$.
It would be interesting to mention that the above range of [Si/H] 
covers the abundance of Si observed for SDSS J102915+172927 
([Si/H] $= -4.27$, Caffau et al.\ 2011).
This could suggest that the growth of Si grains might have worked in 
the parent cloud of this star.

\begin{figure}
\epsscale{1.0}
\plotone{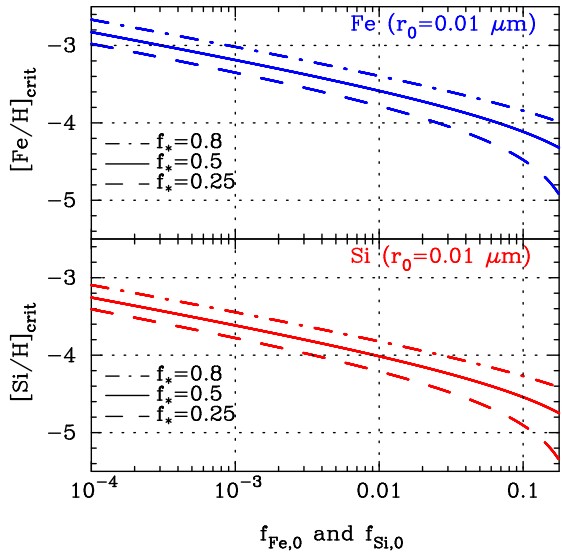}
\caption{
 Critical abundances of Fe (upper panel) and Si (lower panel), for
 which the grain growth can lock up a fraction $f_{i,*}$ = 0.8 
 (dot-dashed), 0.5 (solid), and 0.25 (dashed) of Fe and Si atoms in 
 dust grains at the hydrogen number density $c_{{\rm H},*} = 10^{12}$ 
 cm$^{-3}$.
 The horizontal axis is the initial dust abundance $f_{i,0}$, and the 
 initial grain radius is set to be $r_{i,0} =0.01$ $\mu$m.
\label{fig3}}
\end{figure}

Here we present how the fragmentation condition induced by grain 
growth depends on the unknown parameters such as $r_{i,0}$ and 
$s_i$.
The time duration $\Delta t^{\rm ff}$ for which the gas density 
increases from $c_{\rm H,0}$ to $c_{\rm H,*}$ by free-fall is given 
by $\Delta t^{\rm ff} = 2 \tau_0^{\rm ff} 
(1 - \sqrt{c_{\rm H,0} / c_{\rm H,*}})$.
On the other hand, the time duration $\Delta t^{\rm gg}$ for which 
the condensation efficiency $f_{i,0}$ at $c_{\rm H,0}$ increases up 
to $f_{i,*}$ at $c_{\rm H,*}$ through grain growth is derived from 
Equation (8) as 
$(\Delta t^{\rm gg})^{-1} = (K_i \tau_0^{\rm gg})^{-1}
+ (2 \tau_0^{\rm ff})^{-1}$.
Since $\Delta t^{\rm ff} / \Delta t^{\rm gg} \ge 1$ is required 
for the fragmentation, we can obtain the condition
\begin{eqnarray}
(1.0-2.5)
\left( \frac{s_i}{1.0} \right)
\left( \frac{A_i}{2.5 \times 10^{-9}} \right)
\left( \frac{0.01~\mu{\rm m}}{r_{i,0}} \right) \nonumber \\ 
\times
\left( \frac{c_{{\rm H,*}}}{10^{12}~{\rm cm}^{-3}} 
\right)^{\frac{1}{2}} 
\left( \frac{1.0}{K_i} \right)
\ge 1.0,
\end{eqnarray}
where the numerical factor 1.0 (2.5) is for Fe (Si) grains.
This inequality demonstrates that the fragmentation condition is
achieved more easily for the initial grain radii smaller than
$0.01$ $\mu$m.
In contrast, if $r_{i,0} > 1.0$ $\mu$m or $s_i < 0.01$, the grain 
growth can no longer induce the fragmentation for 
$A_i \la 2.5 \times 10^{-7}$ ([Fe, Si/H] $\la -2$).
We also note that a smaller $f_{i,0}$ producing a larger $K_i$ acts 
against the gas fragmentation by grain growth.

Next we consider the dust-to-gas mass ratio to see whether the 
condition for the dust-induced fragmentation given in Equation (1) can 
be met.
In the context of this paper, we suppose that the product of 
$\mathcal{S}$ and $\mathcal{D}$ without grain growth is given as 
$(\mathcal{S D})_{i,{\rm crit}} = 3 f_{i,0} A_{i,{\rm crit}} m_i / 
4 \rho_i r_{i,0} \mu m_{\rm H}$, adopting the critical metal 
abundance evaluated in Equation (9).
On the other hand, the product resulting from the grain growth 
is given by $(\mathcal{S D})_{i,*} = (\mathcal{S D})_{i,{\rm crit}} 
(f_{i,*} / f_{i,0})^{2/3}$.
Figure 4 shows the dependence of 
$(\mathcal{S D})_{i,{\rm crit}}$ and $(\mathcal{S D})_{i,*}$ on 
$f_{i,0}$, adopting $f_{i,*} = 0.5$.
We can see that $(\mathcal{S D})_{i,{\rm crit}}$ is well below the 
minimum value required for the dust-induced fragmentation 
(dot-dashed line in Fig.\ 4), whereas $(\mathcal{S D})_{i,*}$ exceeds 
this value.
This indicates that, even if the destruction by the SN reverse shock 
results in a lower $\mathcal{S D}$ than the criterion for the 
dust-induced fragmentation, the grain growth can enhance 
$\mathcal{S D}$ in the clouds and can enable the gas fragmentation 
into sub-solar mass clumps.
Note that the results in Figure 4 are independent of the initial grain 
radius $r_{i,0}$ since $\mathcal{S}_i \propto r_{i,0}^{-1}$ and 
$\mathcal{D}_i \propto A_{i,{\rm crit}} \propto r_{i,0}$.

\begin{figure}
\epsscale{1.1}
\plotone{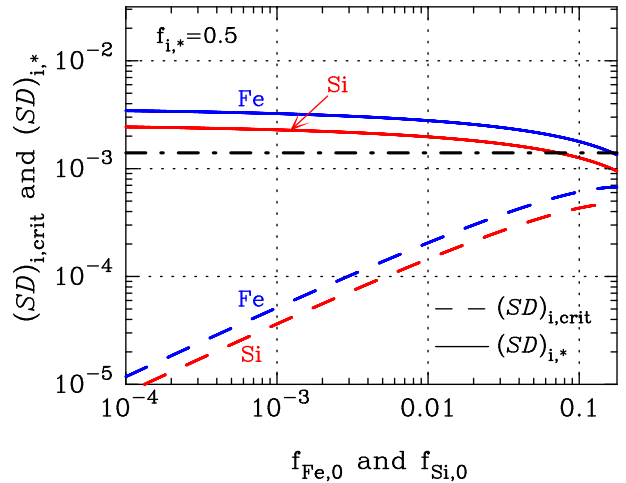}
\caption{
 Products of dust-to-gas mass ratio $\mathcal{D}$ and geometrcal cross 
 section per unit dust mass $\mathcal{S}$ for Fe (blue) and Si (red) as 
 a function of the initial dust abundance $f_{i,0}$.
 The dashed lines depict $(\mathcal{SD})_{i,{\rm crit}}$ corresponding 
 to the critical abundances given by the solid lines in Figure 3, 
 whereas the solid lines depict the resulting dust-to-gas mass ratio 
 after the grain growth $(\mathcal{SD})_{i,*}$.
 The horizontal dot-dashed line indicates the minimum value above which
 the dust emission cooling causes the gas fragmentation into low-mass 
 clumps (Schneider et al.\ 2012a).
\label{fig4}}
\end{figure}

Finally, we relate the critical abundance to the total metallicity $Z$.
Equation (9) suggests that the critical metal abundances are generally 
in the range of $A_{\rm crit} \simeq$ $10^{-9}$--$10^{-8}$, 
depending on $f_{i,0}$, $f_{i,*}$, $r_{i,0}$, and grain species.  
By representing the mass ratio of refractory elements condensible 
into dust grains to the total heavy elements as $\mathcal{R}$, the 
metallicity $Z$ can be related to $A_{\rm crit}$ as
$\mathcal{R} Z \simeq A_{\rm crit} \left( \mu_{\rm ref} / \mu \right)$
with $\mu_{\rm ref}$ being the mean atomic mass of refractory elements.
Then, we have
\begin{eqnarray}
Z \simeq (5-50) \times 10^{-6} \left( \frac{0.2}{\mathcal{R}} 
\right) Z_\odot,
\end{eqnarray}
where we use $\mu_{\rm ref} / \mu =20$ and $Z_\odot = 0.02$.
Equation (11) implies that, if the grain growth does work efficiently, 
it can drive the gas fragmentation of low-mass clumps in the star-forming 
clouds enriched with metallicity $\sim$$10^{-5}$ $Z_\odot$.
In other words, as long as the abundance of a given refractory element 
satisfies the critical abundance in Equation (9), the formation of 
hyper-metal-poor low-mass stars with the metallicity lower than 
$Z \simeq 4.5 \times 10^{-5}$ $Z_\odot$ observed in 
SDSS J102915+172927 could be possible.

\section{Concluding Remarks}

We have investigated the growth of dust grains in metal-poor 
proto-stellar clouds.
Our simple model shows that the grain growth can operate efficiently 
even in collapsing dense cores with metal abundances as low as 
[Fe, Si/H] $\simeq -5$.
We also present the critical metal abundances above which the grain 
growth could affect the fragmentation process of collapsing gas 
clouds.
This abundance is estimated to be 
$A_{\rm crit} \simeq$ $10^{-9}$--$10^{-8}$, which suggests that
the formation of low-mass stars with metallicity of
$\sim$$10^{-5}$ $Z_\odot$ can be possible.
We conclude that even if the initial dust-to-gas mass ratio does not 
satisfy the condition required for the dust-induced fragmentation, 
the grain growth can increase the dust-to-gas mass ratio high enough 
to facilitate the formation of metal-poor low-mass stars.

Our results suggest that if grain growth is considered, the formation
of low-mass protostars can occur not only at very low metallicity but 
also at higher metallicity.
The final mass of a newly born star is determined by the accretion of 
the surrounding gas onto the protostars (McKee \& Ostriker 2007 and 
references therein), and its mass accretion rate would be regulated 
by the mass of the central protostar induced by the grain growth.
Thus, the grain growth in collapsing clouds might be a fundamental 
physical process to control the stellar initial mass function in 
the present universe.
 
It should be mentioned that we have considered only the growth of 
single-component Fe and Si grains with a single initial radius.
However, it might be possible that Si and Fe atoms condense as 
silicates or oxides in an oxygen-rich gas.
Since the growth of such compound grains
with no monomer molecule has been usually treated by considering
Si or Fe element as a key element (e.g., Zhukovska et al.\ 2008),
the mass and radius of dust given in this paper are
considered to be lower limits.
On the other hand, the timescale of grain growth is sensitive to the 
initial grain radius (Hirashita \& Kuo 2011).
Thus, the effect of the initial size distribution as well as the 
growth of compound grains should be explored.
Furthermore, we have assumed the sticking probability of $s_i = 1$ 
and a constant gas temperature $T_{\rm gas} = 10^3$ K during the 
collapse of the clouds.
In particular, too low sticking probabilities ($s_i \la 0.01$) may
prevent the grain growth from becoming efficient for the metal 
abundances of $A_i \la 2.5 \times 10^{-7}$.
We note that our conclusions obtained by the simple model 
should be confirmed by more sophisticated simulations of the 
thermal evolution of star-forming clouds involving grain growth.

\acknowledgments

We thank Hiroyuki Hirashita for useful comments.
We are grateful to the anonymous referee for critical comments
that improved the manuscript.
This research has been supported by World Premier International 
Research Center Initiative (WPI Initiative), MEXT, Japan, and by the 
Grant-in-Aid for Scientific Research of the Japan Society for the 
Promotion of Science (22684004, 23224004).


\end{document}